# A Survey on Self-healing Software System

Zahra Yazdanparast

*School of Electrical and Computer Engineering,
Tarbiat Modares University,
Tehran, Iran*
zahra.yazdanparast@modares.ac.ir

**Abstract**

With the increasing complexity of software systems, it becomes very difficult to install, configure, adjust, and maintain them. As systems become more interconnected and diverse, system architects are less able to predict and design the interaction between components, deferring the handling of these issues to runtime. One of the important problems that occur during execution is system failures, which increase the need for self-healing systems. The main purpose of self-healing is to have an automatic system that can heal itself without human intervention. This system has predefined actions and procedures that are suitable for recovering the system from different failure modes. In this study, different self-healing methods are categorized and a summary of them is presented.

**keywords:** Software engineering, Self-healing, Failure, Software repair.

## 1. Introduction

Testing and reviewing program codes are well-known methods for improving the quality and robustness of software. Unfortunately, the complexity of modern software systems makes it impossible to predict all possible problems that can occur at runtime. This problem makes it impossible to reveal all software errors by using tests and checks. Therefore, there is a need for self-healing software systems [1]. A self-healing system can detect runtime disturbances such as failures and reacts to them by dynamically adapting and reconfiguring the system. Self-healing systems are usually characterized by limitations (e.g., adaptation is required only in case of failure) [2].

The feature of self-healing is dependent on the two features of self-diagnosis and self- healing, and it means the ability to discover and recognize abnormalities and perform appropriate reactions in the face of them. Self-diagnosis means identifying errors, defects or failures in the software system, and self-healing means restoring the system and solving these problems automatically. A system with self-healing capability has the ability to predict potential problems and perform appropriate reactions to prevent system failure [3].

In many studies, the main goals of a self-healing system are defined as maximizing availability, reliability, survivability, and maintainability. In general, the primary source of studies related to the self-healing feature is common methods related to fault tolerance issues, so that these methods can be applied automatically or with minimal human intervention [4, 5].

So far, most studies have been conducted on code-level errors, and fewer studies have been performed on self-healing. In this study, a classification of these methods will be presented.

## 2. Self-healing systems

Figure 1 shows the classification of studies conducted in the field of self-healing. In the following, these categories will be surveyed.

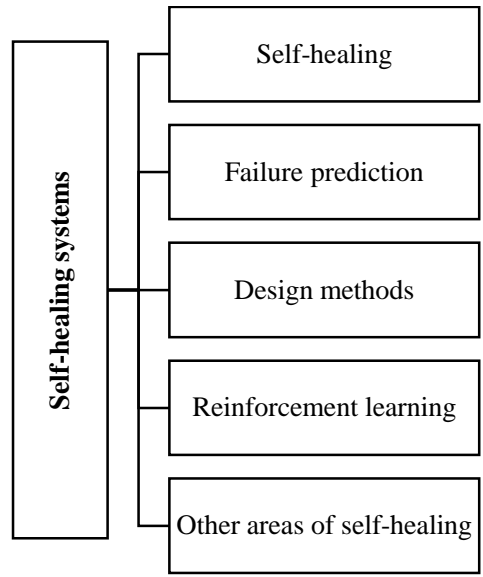

Figure 1: Self-healing articles category.

## 2.1. Self-healing

Vogel et al. [6] proposed a feedback loop-based model for the self-healing process from runtime failure. During this process, the software system is constantly monitored and it is determined whether the system needs to be updated or not based on the system architecture model as well as the triple graph grammar rules (to synchronize the self-adaptation software with the architecture model). If needed, the system is analyzed and if there is a failure in the system, repair decisions are made and finally, these decisions are applied to the system and architectural model.

Aes et al. [7] presented the SHOWA framework that collects the execution time of application transactions and data from the system to detect performance anomalies. The collection of application information is performed by the sensor section, which performs application-level monitoring. Failure scenarios include CPU load, workload, remote service changes, application changes, and energy consumption, which are common causes of failure in web applications. Two simulators TPC-W [8] (retail store website simulation) and RiceUniversity Bidding System (RUBiS) [9] (auction site simulation) have been used for evaluation.

Ghahremani et al. [10] study presented a hybrid rule-based and utility-based method that has the advantages of both approaches. Rule-based approaches define adaptation in advance, and if certain conditions occur, the adaptation is executed. These approaches are extensible but rarely satisfy adaptability decisions. Utility-based approaches identify optimal decisions using costly optimization. These approaches are not scalable to large problems. So that the adaptation decisions are optimal and can be expanded. This approach has been used for self-healing at the architectural level for large software systems. In the proposed method, the impact of each law on the system's usefulness is calculated and for restoration, a law that maximizes the architecture's usefulness is selected.

Angarita et al. [11] presents a knowledge-based approach for self-healing hybrid service applications. A composite service is an application that combines a set of interacting services that call the web. This paper uses agents for failure control and recovery strategy selection. The knowledge used to make decisions is based on precomputed global and local offline information, user QoS preferences, and real-time runtime information.

Albassam et al. [12] presented a model-based approach for self-healing and self-configuration using communication recovery. Therefore, in component-based architecture and service-oriented architecture, a component or service can be dynamically recovered from failure. The design is based on the MAPE-K model. Software recovery involves the dynamic replacement of service, coordinator, and client components after a runtime failure.

Loukil et al. [13] proposed an approach based on runtime models and aspect-oriented software design and implementation to model and execute reconfiguration at runtime. The proposed process covers all stages of the software design and implementation cycle from architectural specification to code generation. The proposed middleware has the possibility of dynamic reconfiguration of the system, monitoring some structural features, maintaining system consistency during reconfiguration, and establishing a causal relationship between the architectural model and the running system.

Combining the functionality of multiple services into a single service mashup has received widespread attention in recent years. Due to the distributed nature of these services, where the manufacturer services are located on different servers, a change in the functionality or availability of the manufacturer service may result in service failure. In Bashari et al. [14] study, a new method based on the software production line engineering (SPLE) pattern is presented, which can find a valid service mashup that has the maximum possible number of main features. The advantage of this method is that it can automatically recover or mitigate the failure without requiring the user to define any rules or adaptation strategies. The content model has been used to create a connection between services and features.

Rajput et al. [15] proposed a new architecture that supports agent-based distributed systems and enables error recovery to achieve self-adaptation. Unlike the traditional multi-agent architecture, it is designed to be self-healing for various task-oriented multi-agent communication activities. The adaptability of the agent's communicative control flow, using three mechanisms such as design, action, and representation, has been proposed as the agent's critical responsibility. The proposed architecture also examines resource failures and availability using performance metrics.

Alhosban et al. [16] presented the SFSS method which is a failure management framework for service-oriented architecture that predicts, detects, and resolves failures. In this method, exception failure handling strategies are identified based on QoS. Several recovery designs have been developed and evaluated based on component service performance. Finally, the best design is implemented.

## 2.2. Failure prediction

A failure can be a software error, performance anomaly, system malfunction, or a combination of these. The Grohmann et al. [17] study has categorized service and architecture level failures based on three criteria: prediction objective, time and model type. Traditionally, to find failures in execution traces, engineers would search for words like "kill", "fail", "error" and "exception".

Lin et al. [18] proposed the LogCluster method. In this method, the execution trace is clustered so that errors can be identified more easily. First, the rejection sequence is converted into a vector. Each rejection event has a different importance, so each event is assigned a weight vector. The amount of similarity between two execution trace sequences is measured and by using cumulative hierarchical clustering technique, similar execution traces are placed in a cluster. Then a representative of the rejection sequence is selected for each cluster. In practice, many of the failures that occur in the online service system are frequent failures. By comparing the execution rejection with the previous knowledge, the occurrence of failure is detected. The proposed method is evaluated on the Microsoft online service system.

Du et al. [19] presented an approach for software failure analysis based on sequence of events. Patterns in the sequence of software events are identified and normal and abnormal behaviors are categorized with the purpose of detecting failure. Sequence classification is difficult, because most classifiers work with feature vectors, while sequence data does not have explicit features. Sequence classification methods are divided into three categories: pattern-based methods, sequence distance-based methods, and model-based methods. Three types of errors are considered:

1. Omission bugs: not calling some methods when called.
2. Additional bugs: injection of additional events causes failure.
3. Ordering bugs: the order of events is wrong.

Borkowski et al. [20] presented an architecture for a Business Process Management System (BPMS). When a process is executed by BPMS, events are collected by event-based systems (EBS). Then the event-based failure prediction (EFP) component analyzes the events. Based on these analyses, the EFP component may predict that failure will occur in the future. At the core of the proposed method, an artificial neural network model is designed, which is responsible for analyzing the input data stream, for failure prediction.

## 2.3. Design methods

The aim of Jamshidi et al. [21] was to reduce the configuration search space in two steps. In the offline phase, machine learning was used to learn the system efficiency model and identify optimal configurations. In the online phase, a design based on quantitative verification was used to select the best adaptation from the set of optimal configurations identified in the offline phase.

Laleh et al. [22] examine the topic of web service composition with a focus on recovering from constraint verification failures. Automated service composition aims to simplify complex tasks by integrating core web services into a workflow to create enhanced services. While many methods for automated web service composition exist, they often rely solely on matching input/output parameters, overlooking applicable terms and usage restrictions set by service providers. To address this, the authors propose a constraint-aware failure recovery approach to anticipate failures within composite services. Their method involves predicting failures and minimizing service returns caused by constraint verification failures through a design-based algorithm and constraint processing technique.

## 2.4. Reinforcement learning

Zhao et al. [23] presented a framework with two capabilities of learning adaptive rules based on goals in the offline phase and evolving adaptive rules based on real-time information about the environment and user goals in the online phase for self-adaptation. Reinforcement learning and case-based Reasoning have been used to create and evaluate rules. It considers two dynamic features, the number of requests per unit and the amount of CPU allocation, which change at runtime. According to dynamic properties, configuration properties take different values and switch between different states. The evaluation is performed on the RUBiS simulator.

Wu et al. [24] applied reinforcement learning based on response time to an online book store. Hrabia et al. [25] proposed a new method for hybrid decision making by adding deep reinforcement learning to ROS Hybrid Behavior Planner (RHBP).

Palm et al. [26] used policy-based reinforcement learning that applies a different type of reinforcement learning and represents the learned knowledge as an artificial neural network. Previous papers used value function in reinforcement learning but this paper uses neural network. It has been checked for workload evaluation in RUBiS simulator.

## 2.5. Other areas of self-healing

Since the design of the self-healing component is complex and expensive, it is not possible to add self-healing to all components. Therefore, Tarinejad et al. [27] identifies the effective components according to their reliability and adds self-healing to only those components. First, a method for modeling self-healing behavior has been proposed using Markov chain. Then, with different combinations of Taylor series and self-healing, several criteria are presented to evaluate the reliability of the software system.

Omar et al. [28] uses big data for self-healing in 5G networks. Using the decision tree and the collected data, the state of each cell is determined, and in case of failure, repair is performed.

Dias et al. [29] performed self-healing in IOT. Broker failure, sensor failure, and connection failure have been checked. Repair strategies include replacement, balance, isolation, persistence, and displacement.

In the study by Aktas et al. [30], CPU, memory, battery and bandwidth failures are considered. In this way, if the value of each exceeds a threshold, failure is detected. It used runtime verification for crash detection.

Magableh et al. [31] presented a microservice architecture to solve the Docker swarm cluster problem in the cloud. The proposed method suggests a self-adaptation feature by following the MAPE-K model. The main contribution of this method is to use the utility function in the adaptation process.

Ratasich et al. [32] presents a framework designed for the self-repair of cyber-physical systems (CPS), which involves detecting failures through data collection and analysis, and enhancing components through design and implementation. This framework incorporates structural adaptability, which involves modifying the system's structure during runtime by adding, removing, or rearranging components, as well as parameter adaptation, which adjusts a component's behavior by altering its parameters. Ontology-based Runtime Reconfiguration (ORR) emphasizes structural adaptation in real-time service-based systems, allowing for the automatic design of replacements when a service failure occurs. ORR employs a knowledge base, or ontology, containing relationships between features in the CPS, along with additional runtime information stored in a table known as Service-to-Ontology Mapping (SOM), to model dynamic changes such as the addition, removal, or adaptation of services during runtime. Unlike many other runtime adaptation techniques, ORR is particularly well-suited for real-time systems.

Seebach et al. [33] has presented a method for self-healing of car cruise system. In this article, it is assumed that it is possible to configure the software at runtime, while in reality; this is not possible due to the configuration of electrical control units (ECUs) functions dynamically at design time. During reconfiguration, the cruise system is disabled and restarted. The self-healing scenario is expressed based on the reallocation of components. We assume that the part of cruise system i.e. ACC-ECU is broken. Therefore, its related components are reassigned to other ECUs. For better efficiency, the related components are placed on an electrical control unit. For faster message exchange and higher throughput, CAN-bus is used instead of bus. The challenge of this method is that it is necessary to turn off the self-driving system and give control to the driver and turn it on again after reassigning the components.

Cloud computing has many problems including failure. Devi and Muthukannan [34] proposed a fault-tolerant self-healing technique that can handle user request execution when a failure occurs. CloudSim has been used for evaluation.

Khalifa et al. [35] proposed a distributed hole detection and repair (DHDR) approach in WSN that performs restoration using only nodes located in the network. The proposed method selects suitable nodes with maximum coverage and minimum energy consumption. These nodes move where they fill the empty area without disrupting communication.

Ochoa-Aday et al. [36] focuses on using self-healing to increase SDN stability. In traditional methods, there are multiple back-up paths, and performing routing calculations is reactive and time-consuming. In the proposed approach, the control paths are quickly recovered by local switch actions and subsequently optimized with the knowledge of the overall controller.

In Gill et al. [36] study, cloud failures are categorized. Most of the articles in this field consider efficiency, resource allocation, workload, and scheduling. Shittu et al. [37] presented a review paper on self-healing strategies for power grid stability.

**3. Conclusion**

Effective debugging is essential to produce software with reliability and high quality. Program debugging is an activity to reduce software maintenance costs. As software complexity increases, failures also increase.

Minimal research has been conducted on the subject of software self-repair at the architectural level. In the field of self-management, most articles and methods consider self-adaptation. Also, the border between self-configuration, self-optimization, and self-healing is not very clear, and sometimes self-healing articles focus on the concepts of optimization and efficiency improvement. Therefore, an attempt was made to provide a summary of articles related to self-healing.